# COSMIC RAY NEON, WOLF-RAYET STARS, AND THE SUPERBUBBLE ORIGIN OF GALACTIC COSMIC RAYS

Short title: Superbubble Origin of Cosmic Rays

W.R. Binns[*], M.E. Wiedenbeck[†], M. Arnould[‡], A.C. Cummings[**], J.S. George[**], S. Goriely[‡], M.H. Israel[*], R.A. Leske[**], R.A. Mewaldt[**], G. Meynet[††], L. M. Scott[*], E.C. Stone[**], and T.T. von Rosenvinge[***]

[*] *Washington University, St. Louis, MO 63130 USA, wrb@wuphys.wustl.edu*
[†] *Jet Propulsion Laboratory, California Institute of Technology, Pasadena, CA 91109 USA*
[‡] *Institut d'Astronomie et d'Astrophysique, U.L.B., Bruxelles, Belgique*
[**] *California Institute of Technology, Pasadena, CA 91125 USA*
[††] *Geneva Observatory, 1290 Sauverny, Switzerland*
[***] *NASA/Goddard Space Flight Center, Greenbelt, MD 20771 USA*

## ABSTRACT

We report the abundances of neon isotopes in the galactic cosmic rays (GCRs) using data from the Cosmic Ray Isotope Spectrometer (CRIS) aboard the Advanced Composition Explorer (ACE). These abundances have been measured for seven energy intervals over the energy range of 84≤E/M≤273 MeV/nucleon. We have derived the $^{22}$Ne/$^{20}$Ne ratio at the cosmic-ray source using the measured $^{21}$Ne, $^{19}$F, and $^{17}$O abundances as "tracers" of secondary production of the neon isotopes. Using this approach, the $^{22}$Ne/$^{20}$Ne abundance ratio that we obtain for the cosmic-ray source is 0.387 ± 0.007 (stat.) ± 0.022 (syst.). This corresponds to an enhancement by a factor of 5.3±0.3 over the $^{22}$Ne/$^{20}$Ne ratio in the solar wind. This cosmic-ray-source $^{22}$Ne/$^{20}$Ne ratio is also significantly larger than that found in anomalous cosmic rays, solar energetic particles, most meteoritic samples of matter, and interplanetary dust particles. We compare our ACE-CRIS data for neon and refractory isotope ratios, and data from other experiments, with recent results from two-component Wolf-Rayet (WR) models. The three largest deviations of GCR isotope ratios from solar-system ratios predicted by these models, $^{12}$C/$^{16}$O, $^{22}$Ne/$^{20}$Ne, and $^{58}$Fe/$^{56}$Fe, are indeed present in the GCRs. In fact, all of the isotope ratios that we have measured are consistent with a GCR source consisting of about 80% material with solar-system composition and about 20% of WR material. Since WR stars are evolutionary products of OB stars, and most OB stars exist in OB associations that form superbubbles, the good agreement of these data with WR models suggests that superbubbles are the likely source of at least a substantial fraction of GCRs.

*Subject headings:* Galactic cosmic rays—Galaxy: stars—Wolf-Rayet: general—Galaxy:abundances—Galaxy: ISM—ISM: abundances—ISM: Galactic Cosmic Rays

## 1. INTRODUCTION

Several experiments have shown that the $^{22}$Ne/$^{20}$Ne ratio at the Galactic Cosmic Ray (GCR) source is substantially greater than that in the solar wind. The solar-system abundances, which are taken here to be equivalent to pre-solar nebula abundances, for most elements and isotopes are represented best by the C1 carbonaceous chondrite meteoritic abundances. However, for most highly volatile elements including the rare gases, the solar wind (SW) is usually assumed to provide the best representation of the composition of the pre-solar nebula (Anders & Grevesse 1989; Lodders 2003). GCR neon isotopic abundances were measured first by a balloon-borne experiment (Maehl et al. 1975), in which the mean mass of the neon isotopes at the source was determined to be considerably larger than for solar-system material. IMP-7 (Garcia-Munoz, Simpson, & Wefel 1979) established that the excess of neutron-rich isotopes was due to a high $^{22}$Ne abundance. Subsequent experiments on ISEE-3 (Wiedenbeck & Greiner 1981; Mewaldt et al. 1980), Voyager (Lukasiak et al. 1994; Webber et al. 1997), Ulysses (Connell & Simpson 1997), and CRRES (DuVernois et al. 1996) confirmed this overabundance with more precise measurements. For the isotopes measured to date, neon is the only element for which such a large difference of isotopic composition at the source, compared to solar-system abundances, has been obtained.

A number of models have been proposed to explain the large $^{22}$Ne/$^{20}$Ne ratio. Woosley and Weaver (1981) suggested that it could be explained by a model in which GCRs originate in a region of the Galaxy with metallicity (fraction of elements heavier than He) greater than that of the Solar System. The yield of neutron-rich isotopes in massive stars is directly proportional to their initial metallicity. If GCRs originate in a region of higher metallicity than the Solar System, then they should have a larger $^{22}$Ne/$^{20}$Ne ratio than the Solar System. Enhanced abundances of the neutron-rich isotopes of Mg and Si are also predicted by this model, but the measured deviations from solar-system abundances are less than would be expected from this mechanism (Connell & Simpson 1997; Wiedenbeck et al. 2003).

Reeves (1978) and Olive and Schramm (1982) suggested that the Sun might have formed in an OB association. They argued that the pre-solar nebula could have been enriched by ejecta from nearby supernovae (SNe) in the association, and they cite meteoritic evidence of the extinct radioactive isotopes $^{26}$Al and $^{107}$Pd, which indicates that the pre-solar nebula must have been enriched by ejecta from at least one SN within ~$10^6$ y (the approximate half-life of those radioisotopes) of the solar-system formation. Massive precursor stars have short lifetimes, ranging from ~$3\times10^6$y to ~$30\times10^6$y for stars with solar metallicity and initial mass 120 M$_\odot$ and 9 M$_\odot$ respectively (Schaller et al. 1992 and Meynet & Maeder 2000). These SNe could have injected large amounts of $^{20}$Ne and other α-particle nuclei (i.e. even-Z nuclei with A/Z=2) into the pre-solar nebula, thus resulting in the $^{22}$Ne/$^{20}$Ne solar-system ratio being anomalously low owing to the large $^{20}$Ne abundance rather than the GCRs possessing an anomalously high ratio. However, as was the case for the supermetallicity model, this scenario should also result in increased $^{25,26}$Mg/ $^{24}$Mg and $^{29,30}$Si/$^{28}$Si ratios in the GCRs that are not observed.

The most widely accepted mechanism for producing the neon ratio excess was first introduced by Cassé and Paul (1982) and was studied more quantitatively by Prantzos et al. (1987). They suggested that the large $^{22}$Ne/$^{20}$Ne ratio in GCRs could be due to Wolf-Rayet (WR) star ejecta mixed with material of solar-system composition. The WC phase (Willis 1999) of WR stars is characterized by the wind enrichment of He-burning products, especially carbon and oxygen. Also, at the beginning of the He-burning phase, $^{22}$Ne is strongly enhanced as a result of $^{14}$N destruction (e.g. Prantzos et al. 1986; Maeder and Meynet 1993) through the α-capture reactions $^{14}$N($\alpha,\gamma$)$^{18}$F($e^+\nu$)$^{18}$O($\alpha,\gamma$)$^{22}$Ne. An excess of the elemental Ne/He ratio in the winds of WC stars has been confirmed observationally (Willis et al. 1997; Dessart et al. 2000). This is consistent with a large $^{22}$Ne excess and gives support to the idea of Cassé and Paul (1982). The high velocity winds that are characteristic of WR stars can inject the surface material into regions where standing shocks, formed by those winds and the winds of the hot, young, precursor OB stars interacting with the interstellar medium (ISM), can pre-accelerate and mix the WR material into the ISM. From a consideration of detailed WR models, Prantzos et al. (1987) concluded that the observed abundance of the neon isotopes could be accounted for if about 2% of the GCR source material (mass fraction) came from WR stars. An important consequence of the WR model is that it also apparently explained why the $^{12}$C/$^{16}$O ratio in GCRs is ~1, while $^{12}$C/$^{16}$O ~0.4 in the solar system. However, Mewaldt et al. (1989) noted two possible problems: The WR admixture did not explain the depletion of $^{14}$N in GCRs and it also leads to the conclusion that a surprisingly large fraction of the heavy elements in GCRs (~25% of those with $6 \leq Z \leq 28$) must come from WR stars.

Maeder and Meynet (1993) suggested a model in which, like the Woosley and Weaver (1981) model, GCRs at Earth come preferentially from regions with metallicity higher than in the Solar System. They suggest that the GCRs observed at Earth come preferentially from nearer the galactic center, and point out that the ratio of the number of WR stars to that of OB stars increases toward the galactic center (Meylan and Maeder 1983; Van der Hucht et al. 1988). They argue that if that ratio were independent of galactocentric radius, there should not be an enhancement of the $^{22}$Ne/$^{20}$Ne ratio resulting from WR material in the source, since the Solar System would have condensed out of a similar mix of material. However, since that gradient is known to exist, a larger fraction of WR products in the GCRs at Earth is expected in their model. They deduce that, near the Sun, ~5% of the mass accelerated in GCRs should have a WR origin to produce the observed $^{22}$Ne/$^{20}$Ne ratio. This scenario should also result in increased $^{25,26}$Mg/$^{24}$Mg and $^{29,30}$Si/$^{28}$Si ratios in the GCRs that are not observed.

Galactic chemical evolution could also contribute similarly to the high GCR $^{22}$Ne/$^{20}$Ne ratio. Since GCRs represent a much more recent sample of matter than solar-system samples (Yanasak et al. 2001; Wiedenbeck et al. 2001a), it would be more fully processed and possess a higher metallicity than the Solar System. The effect on isotope ratios should be very similar to that in the Woosley and Weaver (1981) and Maeder and Meynet (1993) models. However, the expected variation in the $^{22}$Ne/$^{20}$Ne ratio over the $4.5 \times 10^9$ years since the formation of the solar system is expected to be considerably smaller that the observed ratio (Audouze et al. 1981).

Soutoul and Legrain (1999, 2000) have developed a cosmic-ray diffusion model based upon the idea of GCRs coming from the inner galaxy and the gradient in the WR to OB star ratio with galactocentric radius (Maeder and Meynet, 1993). The assumptions in this model are that 1) the cosmic-ray density exhibits a gradient as a function of galactocentric radius and that as one moves toward the galactic center that density increases, 2) the WR star density decreases more rapidly than the O star density with increasing galactocentric radius (Meylan & Maeder 1983; Van der Hucht 2001) and 3) that these WR stars enrich the Galaxy locally in $^{22}$Ne (Meynet & Maeder 1997; Maeder & Meynet 1993). This model predicts a $^{22}$Ne/$^{20}$Ne ratio that has a weak energy dependence, increasing with energy from 1 to 100 GeV/nucleon. This energy dependence results from the WR/OB gradient combined with differing diffusion distances as a function of energy. However, the model predictions do not extend down to the energies sampled by CRIS. The predicted ratio depends strongly on the exact gradient in the WR to OB ratio that is assumed.

Cosmic-ray acceleration and confinement in superbubbles was originally suggested by Kafatos et al. (1981). Streitmatter et al. (1985) showed that the observed energy spectra and anisotropy of cosmic rays were consistent with such a model. Higdon and Lingenfelter (2003) have recently argued that GCRs originate in superbubbles based on the $^{22}$Ne/$^{20}$Ne excess in GCRs. This expands on their initial work, in which they point out that most core-collapse SNe and WR stars occur within superbubbles (Higdon et al. 1998). In their model, ejecta from WR stars and from core-collapse SNe occurring within superbubbles are mixed with ISM material of solar-system composition and accelerated by subsequent SN shocks within the superbubble to provide the bulk of the GCRs. The calculations of Schaller et al. (1992) and Woosley & Weaver (1995) are used to estimate the yields of $^{20}$Ne and $^{22}$Ne from WR stars and core-collapse SN, and from this Higdon & Lingenfelter (2003) estimate that a mass fraction of $(18\pm5)\%$ of WR ejecta plus SN ejecta, mixed with material of solar-system composition, can account for the measured ACE-CRIS $^{22}$Ne/$^{20}$Ne ratio. They conclude that the elevated $^{22}$Ne/$^{20}$Ne ratio is a natural consequence of the superbubble origin of GCRs since most WR stars exist in OB associations.

In this paper we present measurements of the isotopic composition of neon obtained by the Cosmic Ray Isotope Spectrometer (CRIS) instrument (Stone et al. 1998) on the Advanced Composition Explorer (ACE) spacecraft. We have measured the $^{22}$Ne/$^{20}$Ne ratio as a function of energy over the range $84 \leq E/M \leq 273$ MeV/nucleon and derived the $^{22}$Ne/$^{20}$Ne source ratio using our measured $^{21}$Ne, $^{19}$F, and

$^{17}$O abundances as "tracers" of secondary production of the neon isotopes (Stone & Wiedenbeck 1979). This ratio is compared with that obtained from other samples of cosmic matter: meteorites, interplanetary dust particles (IDP), solar wind (SW), anomalous cosmic rays (ACR), and solar energetic particles (SEP). We then compare the CRIS measurements of neon and heavier refractive isotopic ratios, and measurements from other experiments, with predictions of recent WR models. These WR modeling calculations have been performed by the coauthors from the Institut d'Astronomie et d'Astrophysique, Brussels, and from the Geneva Observatory, Switzerland. We then consider these results in the context of a possible superbubble origin of GCRs (Higdon & Lingenfelter 2003).

## 2. MEASUREMENTS

The CRIS instrument (Stone et al. 1998) consists of four stacks of silicon solid-state detectors to measure dE/dx and total energy ($E_{tot}$), and a scintillating-fiber hodoscope to measure trajectory. The dE/dx-$E_{tot}$ method is used to determine particle charge and mass. The geometrical factor of the instrument is ~250 cm$^2$sr and the total vertical thickness of silicon for which particle mass can be determined is 4.2 cm. The precision with which angle is measured by the fiber hodoscope is ≤ 0.1°.

Figure 1 shows the CRIS neon data in 7 range bins. A histogram of the sum of all events is shown in Figure 2a. These data were collected from 1997 December 5 through 1999 September 24 and are a selected, high-resolution data set. This analysis includes events with trajectory angles ≤25º relative to the normal to the detector surfaces. Particles stopping within 750 μm of the single surface of each silicon wafer having a significant dead layer were excluded from this analysis. Nuclei that interacted in CRIS were identified and rejected by requiring no signal in the bottom silicon anticoincidence detector, requiring consistency in charge estimates obtained using different combinations of silicon detectors for events penetrating beyond the second detector in the silicon stack, and by rejecting particles with trajectories that exit through the side of a silicon stack. The average mass resolution for neon is 0.15 amu (rms). This resolution is sufficiently good that there is only a slight overlap of the particle distributions for adjacent masses. In Figure 1, the total number of neon events is ~4.6 × 10$^4$. With the statistical accuracy of the CRIS data, it is possible, for the first time, to study the energy dependence of the $^{22}$Ne/$^{20}$Ne ratio with high precision.

Table 1 gives the numbers of events by "detector range" in the telescope for $^{20}$Ne, $^{21}$Ne, and $^{22}$Ne and the $^{22}$Ne/$^{20}$Ne and $^{21}$Ne/$^{20}$Ne ratios obtained by fitting data in each detector range, and all ranges summed together. For each detector range the median energy and energy interval of the events detected for that detector range (the energy range encompasses 95% of the particles for that detector range) are calculated and shown in columns 2 and 3. The numbers of events for each of the neon isotopes are listed in columns 4-6 to indicate the statistical accuracy of those measurements. With the high statistical accuracy of these observations it is necessary to also assess the systematic uncertainties associated with the analysis, including the cuts used to select the final data set and the procedure used for deriving relative abundances from mass distributions such as those shown in Figure 1. For this purpose we performed two semi-independent analyses of the neon data and compared the results. The derived isotope ratios were found to agree to within better than 2% rms, which, though small, is still larger than our statistical uncertainties. We assigned a 2% systematic error to the measured ratios, which encompasses both the analysis systematics and the possible effect of uncorrected fragmentation production of the neon isotopes in the instrument, which should be <0.5%. To obtain the final ratios listed in columns 7 and 8, it was necessary to correct for interactions in the instrument. These correction factors, which were obtained using the Westfall (1979) cross-sections, range from 0.2 to 1.2% for $^{22}$Ne/$^{20}$Ne and 0.1 to 0.6% for $^{21}$Ne/$^{20}$Ne. The numbers of events for each isotope are binned in equal range intervals, extending from the minimum to the maximum range (energy) within each detector. However, we want to obtain isotope ratios over equal energy intervals. Since different isotopes of the same element have slightly different energy intervals for a given range interval, it is necessary to calculate an adjustment factor for the ratios. Range-energy adjustment factors range from 6.9 to 8.3% for $^{22}$Ne/$^{20}$Ne and 3.5 to 4.2% for $^{21}$Ne/$^{20}$Ne. The final corrected ratios are given in Table 1 and are plotted in Figure 3 as a function of energy. The uncertainties quoted for the individual detector ranges are statistical only. The GCR $^{22}$Ne/$^{20}$Ne ratio is approximately constant with a small increase with energy. A similar behavior is observed for $^{21}$Ne/$^{20}$Ne. Least-squares linear fits to these ratios, which are shown as dotted lines in Figure 3, are described by the following equations:

$$^{22}Ne/^{20}Ne = (0.584) \times [(1\pm0.010)+(6.58\pm2.24)\times10^{-4}\times(E/M-180 MeV/nucleon)]$$
$$^{21}Ne/^{20}Ne = (0.214) \times [(1\pm0.016)+(8.34\pm3.33)\times10^{-4}\times(E/M-180 MeV/nucleon)]$$

where E/M is in units of MeV/nucleon, and 180 MeV/nucleon is the reference energy used to calculate the source abundances of these ratios.

| Detector Range* | Energy for $^{20}$Ne (MeV/nuc) | | Number of Events | | | Final Corrected Ratio | Final Corrected Ratio |
|---|---|---|---|---|---|---|---|
| | Median | Interval | $^{20}$Ne | $^{21}$Ne | $^{22}$Ne | $^{22}$Ne/$^{20}$Ne | $^{21}$Ne/$^{20}$Ne |
| R2 | 90 | 84-97 | 2583 | 503 | 1381 | 0.5644±0.0188 | 0.2033±0.0099 |
| R3 | 120 | 107-135 | 5029 | 962 | 2624 | 0.5516±0.0128 | 0.2021±0.0068 |
| R4 | 154 | 142-168 | 4103 | 837 | 2293 | 0.5898±0.0133 | 0.2064±0.0072 |
| R5 | 184 | 173-198 | 3533 | 709 | 1953 | 0.5792±0.0141 | 0.2153±0.0080 |
| R6 | 211 | 200-225 | 2883 | 659 | 1705 | 0.6036±0.0152 | 0.2285±0.0089 |
| R7 | 236 | 225-249 | 2568 | 577 | 1462 | 0.5969±0.0163 | 0.2296±0.0095 |
| R8 | 259 | 248-273 | 2254 | 489 | 1322 | 0.6131±0.0192 | 0.2146±0.0102 |
| R-All | | 84-273 | 22954 | 4736 | 12740 | 0.5839±0.0060(stat)±0.012(sys) | 0.2137±0.0033(stat)±0.0043(sys) |

*Range designations (column 1) refer to the sample of particles stopping in the Nth detector. e.g. R2 refers to particles stopping in the detector designated as E2 in Figure 11 of Stone et al. (1998).

Table 1—Numbers of events and isotope ratios.

The CRIS Ne isotopic ratios are also compared with measurements made by other experiments in Figure 3. The measured ratios reported from those experiments (Wiedenbeck & Greiner 1981 [ISEE-3]; Webber et al. 1997 [Voyager]; Connell & Simpson 1997 [Ulysses]; DuVernois et al. 1996 [CRRES]) are plotted as open symbols. The agreement with the other experiments is generally good. The mean levels of modulation for the Voyager and Ulysses measurements were similar to that of our CRIS measurement, although the Voyager data were taken over a wide range of modulation levels. Although ISEE-3 and CRRES measurements were at a different modulation, the effect of that difference is within the statistical uncertainty of their measurements. Therefore the ratios from these experiments have not been adjusted for differing solar modulation. The energy range corresponding to each of the experiments is shown as a horizontal bar at the bottom of Figure 3. The CRIS measurements have sufficient statistics to obtain, for the first time, energy spectra of the neon isotopes over this energy range.

## 3. SOURCE COMPOSITION

To obtain the $^{22}$Ne/$^{20}$Ne abundance ratio at the comic-ray source from the ratio that we have observed, we must account for the secondary contributions to the observed fluxes of these isotopes, i.e., the production of these isotopes resulting from fragmentation of heavier nuclei due to nuclear interactions as they propagate through the interstellar medium. The "tracer method" of Stone and Wiedenbeck (1979) uses observed abundances of isotopes that are almost entirely secondary to infer the secondary contribution to isotopes like $^{22}$Ne, for which the observed fluxes are a mixture of primary and secondary nuclei. $^{21}$Ne is such a "tracer" isotope; the $^{21}$Ne/$^{20}$Ne ratio in the cosmic rays is two orders of magnitude greater than in the solar wind, so the observed ratio is almost entirely due to secondary production of $^{21}$Ne.

We use a leaky-box cosmic-ray propagation model with assumed nominal source abundances and a given escape mean free path $\Lambda$ to calculate the interstellar spectrum for each nuclide. Because cross-section uncertainties introduce uncertainties in these spectra, we adjust the spectrum for each nuclide except $^{20}$Ne, $^{21}$Ne, and $^{22}$Ne with an energy-independent factor so that it agrees with the observed spectrum when solar modulation (modulation parameter $\phi=400\pm60$ MV) is included. Holding these adjusted spectra fixed, we then calculate the interstellar spectra for $^{20}$Ne, $^{21}$Ne, and $^{22}$Ne, iterating the source abundances of $^{20}$Ne and $^{22}$Ne until the model reproduces the observed intensities of $^{22}$Ne and $^{20}$Ne. The $^{21}$Ne/$^{20}$Ne ratio from this iteration can then be compared with the observed ratio at 1 AU to determine if the assumed mean free path results in the correct secondary production.

In our previous estimates of the $^{22}$Ne/$^{20}$Ne ratio (Binns et al. 2001), we used the cross-sections of Silberberg et al. (1998), scaled to cross-sections measured using beams of energetic heavy ions in cases where they were available. In the current analysis we have also used cross-sections measured using energetic protons on Mg and Si targets to further constrain some of the most important reactions for the propagation. The method used to obtain these cross-sections is described in Appendix A and the cross-section values used are listed. For all other reactions, the scaled Silberberg et al. (1998) cross-sections used previously were employed.

In Figure 4a, the solid diagonal line shows how the inferred source abundance ratio of $^{22}$Ne/$^{20}$Ne and secondary ratio of $^{21}$Ne/$^{20}$Ne at 1 AU are correlated as the escape mean free path varies. The escape mean free path dependence on particle rigidity and velocity from Equation 1 of Davis et al. 2000 was used, and the overall coefficient $\Lambda$ (29.5 g/cm$^2$ in Davis et al.) was adjusted to vary the secondary production. The filled point on this line is the result we obtain using $\Lambda=25$ g/cm$^2$, which corresponds to an escape mean free path of 8.44 g/cm$^2$ for a 400 MeV/nucleon ion with A/Z=2. The open circles correspond to 5g/cm$^2$ increments in $\Lambda$ or 1.69 g/cm$^2$ in the mean free path for 400 MeV/nucleon and A/Z=2. Note that larger $\Lambda$ results in more $^{21}$Ne and more secondary contribution to $^{22}$Ne, so the source abundance of $^{22}$Ne decreases with increasing $\Lambda$. The dot-dashed lines parallel to the solid line show how the correlation between the source ratios and the observed trace abundance change when the observed $^{22}$Ne abundance is varied by its $\pm1\sigma$ statistical uncertainty. Similar lines, (not shown) are calculated to correspond to $\pm1\sigma$ variations of the observed $^{20}$Ne. The horizontal dotted lines are the measured $^{21}$Ne/$^{20}$Ne ratio (center line) and the corresponding 1-$\sigma$ measurement statistical uncertainty (top and bottom lines). Vertical dotted lines are drawn at the intersection of these horizontal lines and the solid diagonal line. The intersection of the center vertical dotted line with the abscissa is the best estimate of the $^{22}$Ne/$^{20}$Ne ratio at the GCR source inferred from the $^{21}$Ne tracer, and the right and left vertical dotted lines are the corresponding 1-$\sigma$ uncertainties due to the uncertainty in the $^{21}$Ne/$^{20}$Ne ratio measurement. This analysis, using $^{21}$Ne as the tracer, results in a source ratio for $^{22}$Ne/$^{20}$Ne of 0.3793 $\pm$ 0.0024.

Since the tracer isotope, $^{21}$Ne, is so close in mass to the isotope of interest, $^{22}$Ne, this calculation of the $^{22}$Ne/$^{20}$Ne source ratio is quite insensitive to the details of the cosmic-ray propagation model. The fact that the true propagation is not a simple leaky box should not seriously affect the result. However, the tracer method, like any propagation, depends upon the fragmentation cross-sections that are used. As an estimate of the sensitivity of the result to these cross-sections, we have done two other tracer calculations using $^{19}$F (Figure 4b) and $^{17}$O (Figure 4c) as the tracers rather than $^{21}$Ne. Histograms of these isotopes are shown in Figures 2b and 2c. These calculations give source ratios for $^{22}$Ne/$^{20}$Ne of 0.3899 $\pm$ 0.0025 and 0.3919 $\pm$ 0.0028 respectively. These uncertainties are statistical only. The arithmetic mean of these three source ratios for $^{22}$Ne/$^{20}$Ne is 0.387 and is shown in Figure 4 as the vertical dashed line.

The root-mean-square standard deviation of these three values is 0.0068, and is one source of systematic uncertainty in our results. Other contributions to our estimated systematic uncertainty are cross-section uncertainties (0.016), an uncertainty based on the difference of the two semi-independent analysis methods used combined with possible uncorrected fragmentation production in the instrument (0.014), and the uncertainty in the solar modulation level (0.005).

To estimate the uncertainties associated with the secondary production of $^{20}$Ne and $^{22}$Ne, we used the cross-section uncertainties in Appendix A for production from $^{24}$Mg and $^{28}$Si, the major contributors to secondary production of the neon isotopes. For neon production from other nuclei, a 25% cross-section uncertainty was assumed for reactions for which some measurements exist, and a 50% uncertainty for reactions with no data.  For neon production on helium, all cross-sections were assigned a 50% uncertainty.

Adding these systematic errors in quadrature, we estimate our combined systematic uncertainty to be 0.022. Thus our CRIS value for the $^{22}$Ne/$^{20}$Ne source ratio is 0.387 ± 0.007 (stat.) ± 0.022 (syst.). Table 2 summarizes these measured ratios and uncertainties.

Note that in Figure 4 the scale on the top horizontal axis is the $^{22}$Ne/$^{20}$Ne GCRS ratio relative to the solar wind (SW) ratio which is 0.0730 ± 0.0016  (Geiss, 1973). The combined measurements using the three tracer isotopes, and adding the statistical and systematic uncertainties quadratically, result in a ($^{22}$Ne/$^{20}$Ne)$_{GCRS}$/($^{22}$Ne/$^{20}$Ne)$_{SW}$ ratio of 5.3 ± 0.3.

|  | Tracer Isotopes | | |
|---|---|---|---|
|  | $^{21}$Ne | $^{19}$F | $^{17}$O |
| Source $^{22}$Ne/$^{20}$Ne ratio estimate | 0.3793 | 0.3899 | 0.3919 |
| Uncertainty from tracer statistics | 0.0024 | 0.0025 | 0.0028 |
| Mean of three tracer ratios | 0.3870 | | |
| Statistical Uncertainty | 0.0075 | | |
| Systematic Uncertainty | 0.0220 | | |
| Final Source $^{22}$Ne/$^{20}$Ne ratio estimate | 0.387 ± 0.007 (stat.) ± 0.022 (syst.). | | |

Table 2—Summary of tracer $^{22}$Ne/$^{20}$Ne ratio and uncertainty estimates.

## 4. DISCUSSION

### 4.1 Comparison of Data with "Cosmic" Samples of Matter

Figure 5 compares our result for the $^{22}$Ne/$^{20}$Ne abundance ratio at the cosmic-ray source, (0.387 ± 0.007 (stat.) ± 0.022 (syst.)), with the ratio of these isotopes in other samples of cosmic matter.

*Solar Wind*--The solar wind is generally believed to give the best estimate of most isotopic ratios of noble gases, including the neon isotopes, in the presolar nebula (Anders & Grevesse 1989; Lodders 2003) since their isotopic abundances are thought to undergo less fractionation in the solar wind than in most other samples of matter. It is a sample of material from the solar corona, and its isotopic composition is reasonably stable on time scales comparable to the solar cycle. The solar wind $^{22}$Ne/$^{20}$Ne ratio plotted in Figure 5 is taken from Geiss (1973).

*Solar Energetic Particles*—It is believed that SEPs in gradual events are a sample of the outer corona and that SEPs in impulsive events sample the lower corona (Cohen et al. 2000). The data point indicates the ratio deduced for the solar coronal abundances from SEP observations after empirically accounting for the observed fractionation (Leske et al. 2003). The average is close to the solar wind value.

*Anomalous Cosmic Rays*—Neutral atoms in the very-local interstellar medium drifting into the heliosphere, where they are ionized by solar UV or charge exchange with the solar wind, result in singly ionized "pickup ions".  These ions are then swept out by the solar wind to the outer heliosphere where they are accelerated as ACRs with energies of typically tens of MeV/nuc. The $^{22}$Ne/$^{20}$Ne ratio plotted for ACRs is taken from Leske et al. (1999) and shows good agreement with the solar wind.

*Meteorites and Lunar Samples*—The Ne-A component in meteorites is found in carbonaceous chondrites, and in the past it was believed to be presolar since a substantial fraction of it is carried by presolar diamonds within those meteorites (Huss & Lewis 1994). However, it is now thought to be a mix of presolar and other components (Ott 2002). Ne-B has two components that likely result from solar wind (Geiss 1973) and very low-energy (<0.1 MeV/nucleon) "solar energetic particle" (SEP) implantation in lunar grains (Wieler et al. 1986; Wieler 1998). (We note here that "solar energetic particle (SEP)" as used in meteoritic studies differs from that in solar physics. In solar physics it refers to particles with energies that can extend from tens of keV/nucleon to hundreds of MeV/nucleon). The identification of Ne-B is based on measurements that show that it has a $^{22}$Ne/$^{20}$Ne ratio very similar to that of contemporary measurements of the solar wind. In addition, these components are found very close to the grain surfaces (depths of up to several tens of nanometers and 30 micrometers respectively for the SW and the so called "SEP" components), indicating an implantation origin. Ne-C has a $^{22}$Ne/$^{20}$Ne ratio that is only slightly larger than for Ne-B. However, it is distinct in that, instead of being found very near the surface of grains, it extends in from the grain surface by as much as several millimeters (Wieler et al. 1986). These have been designated as "solar flare" (SF) particles (Black, 1983). Mewaldt et al. (2001) have shown that the "suprathermal tail" component of the solar wind has energies consistent with the Ne-C penetration depths in meteorites. However the ratio of the amount of Ne-C to Ne-B is considerably larger than can be accounted for by the present-day relative flux of suprathermal particles to that of the solar wind (Wimmer-Schweingruber et al. 2001). Ne-E, which is found in some SiC and graphite grains, is nearly pure $^{22}$Ne (Ozima & Podosek 1983). There are two distinct components of Ne-E, designated H and L (for high and low temperature release, respectively). Ne-E(H) is found in SiC grains within meteorites and it is argued that it originates in He-burning asymptotic giant branch (AGB) star envelopes (Lewis, Amari, & Anders 1994). Ne-E(L), on the other hand, is found in graphite grains, and is thought to originate primarily from the decay of $^{22}$Na produced in novae (Clayton & Hoyle 1976; Amari et al. 1995). On the basis of the high $^{22}$Ne/$^{20}$Ne ratio alone, an obvious possibility for the source of Ne-E is WR stars. However, a comparison of the results of modeling calculations for isotopic ratios of carbon, oxygen, aluminum, and silicon with measured meteorite composition has not yielded any meteoritic grains to date that can be unambiguously identified as originating from WR stars (Arnould et al. 1997). Ne-E is the only population found in meteorites that has a $^{22}$Ne/$^{20}$Ne ratio greater than that found for GCRs.

*Interplanetary Dust Particles*—It is believed that the sources of IDP's are from the asteroid belt and cometary dust. As such, interplanetary dust represents primitive matter from both the inner and outer solar system and contains implanted neon that is presumed to be a mix of two distinct components: solar wind and SEPs (in the meteoritic sense of SEPs described above). Kehm (2000) has measured the $^{22}$Ne/$^{20}$Ne ratio for 29 of these grains. In Figure 5, the mean value for 27 of these particles is plotted as a single data point with the solid horizontal bar giving the range of ratios for those grains. (The IDP bar in this figure does not include data from two grains that had very high ratios for which the measurements are in question [Kehm 2000]).

*Galactic Cosmic Ray Source*--The GCR source $^{22}$Ne/$^{20}$Ne abundance ratio that we have obtained, 0.387 ± 0.007 (stat.) ± 0.022 (syst.), is 5.3 ± 0.3 times greater than the solar wind value of 0.073. Enhancement factors in the $^{22}$Ne/$^{20}$Ne GCR source abundance ratio over the solar-system abundance ratio quoted for other experiments (see Section 1) range from ~3 to ~5. We note that the ISEE-3 (Wiedenbeck and Greiner 1981) and IMP-7 (Garcia-Munoz et al. 1979) reports used the Ne-A abundance as the solar-system reference abundance instead of the solar wind value, which results in a lower estimate of this GCR to solar-system ratio. If they had used the SW value as their reference, their source ratios would have been multiplied by (0.12/0.073)=1.64 and their reported source abundances relative to the reference would also have been close to 5. The large difference in the $^{22}$Ne/$^{20}$Ne ratio in GCRs compared with that in ACR's is striking. This deviation is also seen directly in ACE-SIS (Leske et al. 1999) and SAMPEX results (Leske et al. 1996). The ACRs sample the very local ISM and are apparently not representative of the source material for GCRs.

**4.2 Superbubble and Wolf-Rayet Models of Galactic Cosmic Ray Origin**

Supernovae shocks are believed to be the accelerators of GCRs for energies $<\sim 10^{15}$ eV. By far the majority of core-collapse SNe in our galaxy (~90%) are believed to occur in OB associations that form superbubbles within giant molecular clouds (Higdon et al. 1998; Higdon & Lingenfelter, 2003). Likewise, most WR stars are observed in OB associations and many of their O and B star constituents are expected to transform into WR's in the course of their evolution (Knödlseder et al. 2002; Maeder, 2000). Van Marle, et al. (2005) have performed 2-dimensional modeling of 35-$M_\odot$ OB stars in ISM from star formation through the WR phase. This model shows that winds of the hot, young, OB stars blow bubbles in the ISM with radius ~40 pc. This is followed by a burst of high velocity winds when the star enters the WR phase, and finally the star undergoes a core-collapse SN. The lifetime of these massive stars is short, typically a few million years, and the WR phase is typically a few hundred thousand years (Meynet & Maeder 2003). It therefore seems almost certain that pre-supernova WR wind material will be swept up and accelerated either by the SN shock from the evolved WR star that ejected the material in the first place or by nearby SNe resulting from short-lived massive O and B stars, without substantial mixing into the ambient ISM outside the superbubble. It has been estimated that typically $10^{-4}$-$10^{-5}$ solar masses of material per year is ejected from individual WR stars in high velocity winds (Nugis and Lamers 2000). There are two dominant successive phases of WR stars, the WN and WC phases (Maeder and Meynet 1993). Large quantities of He-burning material rich in $^{22}$Ne are expelled from the stars when they are in the WC phase, resulting in $^{22}$Ne/$^{20}$Ne ratios in the wind material that are enhanced by about two orders of magnitude over solar-system abundances. In the WN phase, CNO processed material is ejected with the resultant production of high $^{13}$C/$^{12}$C and $^{14}$N/$^{16}$O ratios, but no significant increase in the $^{22}$Ne/$^{20}$Ne ratio (Prantzos et al. 1987; Maeder & Meynet 1993).

Higdon and Lingenfelter (2003) have calculated the mass of the neon isotopes synthesized and ejected in superbubbles by massive stars in their WR and core-collapse SN phases, and then modeled the mean $^{22}$Ne/$^{20}$Ne ratio within the superbubble as a function of the mixing fraction with old ISM taken from Anders and Grevesse (1989) adjusted for the present-day ISM metallicity. They have used the results of Schaller et al. (1992) to estimate the mass of $^{22}$Ne and $^{20}$Ne ejected by WR stars over their lifetime, and the results of Woosley & Weaver (1995) to estimate the ejecta yields from core-collapse SNe (SNII and SNIb,c). They assumed that the superbubble metallicity is that of the present-day local ISM, $Z_{ISM}$=0.0264 (however, see Lodders 2003 & Asplund et al. 2004 for revised estimates of solar metallicity), and interpolated the Schaller et al. and Woosley and Weaver results to obtain corresponding mass yields. They estimate that a mass fraction, $f_{ej}$=(18±5)%, of WR plus SN ejecta must be mixed with ISM material of solar-system composition in the superbubble core in order to obtain the $^{22}$Ne/$^{20}$Ne ratio that we reported in a preliminary analysis of the CRIS results (Binns et al. 2001), which is is very close to the final results reported here ($f_{ej}$ is defined as the mass fraction of WR plus SN ejecta summed over all nuclei with charge (Z) ≥ 1). It should be noted that most of the $^{22}$Ne comes from the WR outflows, not the SN ejecta. Higdon and Lingenfelter conclude that "the $^{22}$Ne abundance in the GCRs is not anomalous but is a natural consequence of the superbubble origin of GCRs in which the bulk of GCRs are accelerated by SN shocks in the high-metallicity, WR wind and SN ejecta enriched, interiors of superbubbles". They further assert that the measured value of the $^{22}$Ne/$^{20}$Ne ratio provides evidence for a superbubble origin of GCRs.

As a further test of the superbubble model of cosmic-ray origin, we examine other isotope ratios at the cosmic-ray source, inferred from our CRIS observations and others. These ratios are compared with modeling calculations of WR outflow presented below that provide predictions of isotope ratios in addition to $^{22}$Ne/$^{20}$Ne. It should be noted that these model results do not include explicit core-collapse

SN ejecta contributions to those ratios as was the case for the Higdon and Lingenfelter work on the neon isotopes described above.

We use here the recent massive-star models with metallicity Z=0.02 of Meynet and Maeder (2003) with rotational equatorial velocities at the surface on the Zero Age Main Sequence of either 0 or 300 km/s. A detailed description of the physics of these models can be found in the above reference. Additionally, calculations have been performed for other metallicities, but these are not considered in this paper. The stellar models follow the evolution of the main nuclear species up to $^{26}$Mg. The models with rotation are consistent with the observed number ratio of WR to O-type stars in the solar neighborhood. Additionally they are consistent with the observed ratio of type Ib/c to type II SNe, and for the existence of a small, but observable, fraction of WR stars with both H and He-burning products at their surface. This good agreement of the modeling results with observations is achieved not only for the rotating Z=0.02 models but also for different metallicities (Meynet and Maeder 2005). While these results do not pertain directly to the comparisons that we will make below, they do provide an independent validation of the rotating stellar models. The non-rotating models, on the other hand, have difficulties in reproducing the above observational constraints.

Based on the physical conditions derived from these models, an extended nuclear reaction network is solved in order to follow the evolution of the abundances in the WR winds of the nuclides in the whole $6 \leq Z \leq 82$ range. (Note that the inclusion of these additional reactions does not affect the energetics based on the reduced network used to model the stars). On such grounds, we calculate for each model star the amounts of each nuclide ejected in the wind between the Zero Age Main Sequence and the end of the WR phase. The material made of this mixture is referred to as the WR wind material.

For each WR model star, we find what mixture of WR outflow with material of solar-system (solar-wind) composition would give the $^{22}$Ne/$^{20}$Ne ratio found by CRIS for the GCR source. Table 2 shows the fraction (p) of WR material thus required for each case. The mixing fractions shown here are considerably larger than in the earlier work (Meynet et al. 2001). This difference is largely due to the use of the CRIS $^{22}$Ne/$^{20}$Ne ratio of 5.3 relative to the Solar System instead of 3.0 used in that earlier work; it also reflects the use of lower mass-loss rates from WR stars (Nugis and Lamers 2000) and the use of an updated reaction network and reaction rates. The especially large p values derived for the rotating 85 and 120 M$_\odot$ stars relate to the fact that they have a very long WN phase. Consequently, they lose enormous amounts of CNO processed material that is not $^{22}$Ne-enriched. Except for these two cases, the mixing fractions in Table 2 are similar to the value of 0.18±0.05 derived by Higdon and Lingenfelter. The high p-values predicted for the M ≥ 85 M$_\odot$ stars are not a problem, however, since these very massive stars are expected to be much rarer than the lower mass ones if one adopts a Salpeter-type Initial Mass Function (IMF) (Salpeter 1955), which predicts that the number of stars born at each time with an initial mass M is proportional to M$^{-2.35}$.

| WR Initial Mass (M$_\odot$) | No Rotation WR Fraction (p) | Rotation WR Fraction (p) |
|---|---|---|
| 40 | --- | 0.22 |
| 60 | 0.20 | 0.16 |
| 85 | 0.12 | 0.44 |
| 120 | 0.16 | 0.37 |

Table 2—The mass fraction of ejecta from WR stars, integrated from the time of star formation, mixed with material of solar-system composition that is required to normalize

each model to the CRIS $^{22}$Ne/$^{20}$Ne ratio. The non-rotating model predicts that a 40 M$_\odot$ initial mass star does not go through the WR phase.

The ratios of other isotopes that result from the WR mix with material of solar-system composition that is required to match the $^{22}$Ne/$^{20}$Ne ratio are shown in Figure 6 for non-rotating and rotating models respectively. The results of these models are compared with the GCR source ratios inferred from CRIS and other observations. The plotted neon point (closed circle) is the source ratio inferred from our propagation described above. The points for heavier elements are also from CRIS results (Wiedenbeck et al. 2001a and 2003). Ulysses Mg and Si data (Connell and Simpson 1997) are in good agreement with our CRIS results, while their $^{58}$Fe/$^{56}$Fe ratio (Connell 2001) is significantly lower than the CRIS value. A possible reason for this discrepancy for Fe has been suggested by Wiedenbeck et al. (2001b). We have not plotted their data point since the error bars are large. The solid diamonds plotted for the lighter elements are mean values of GCR source abundances, divided by the Lodders (2003) solar-system abundances, and weighted by their published uncertainties, obtained from Ulysses (Connell and Simpson 1997), ISEE-3 (Krombel and Wiedenbeck 1988; Wiedenbeck and Greiner 1981), Voyager (Lukasiak et al. 1994) and HEAO-C2 (Engelmann et al. 1990). The plotted error bars are weighted means from these experiments. The mean values are obtained from these experiments as follows: $^{12}$C/$^{16}$O—Ulysses and HEAO-C2 (note that these are actually element ratios that have not been corrected for the small fraction of neutron-rich C and O isotopes present at the source); $^{14}$N/$^{16}$O—ISEE-3, Voyager, and HEAO-C2; N/Ne—Ulysses and HEAO-C2. All ratios plotted here are relative to the Lodders (2003) solar-system abundances.

For nuclei heavier than neon, we see that the WR models are in reasonable agreement with data (within about 1.5 sigma), with the exception of the high-mass (85 and 120 solar masses) rotating star models that predict a deficiency in the $^{25}$Mg/$^{24}$Mg ratio, which is not observed. The observed enhancement of $^{58}$Fe/$^{56}$Fe is roughly consistent with the enhancement of this ratio predicted by the models. The GCR data do not show any significant enhancement of the $^{26}$Mg/$^{24}$Mg, while the models do show some enhancement. The difference is less than 1.5 standard deviations. Moreover, the cross-section used in the WR models for the reaction $^{22}$Ne($\alpha,\gamma$)$^{26}$Mg is uncertain (Angulo et al. 1999), and decreasing this cross-section within the range of its uncertainty significantly reduces the predicted $^{26}$Mg/$^{24}$Mg ratio, bringing it into agreement with the GCR result.

Each of the ratios compared for nuclei heavier than neon are for measured isotopes of the same element. For elements lighter than neon, there is generally only a single isotope for which source abundances can be obtained with precision sufficient to constrain the models. Therefore the ratios compared are for different elements. This makes comparisons more complicated since atomic fractionation effects may be important for some ratios. If we compare the plotted data for nuclei lighter than neon with modeling predictions, initially ignoring elemental fractionation effects, we see in Figures 6a and 6b that the measured $^{12}$C/$^{16}$O source ratio is much larger than in the Solar System and is in qualitative agreement with the WR models for non-rotating stars, and rotating stars with initial masses of 40 and 60 M$_\odot$. It is in strong disagreement with models with rotating initial mass stars of 85 M$_\odot$ and 120 M$_\odot$.

The experimental $^{14}$N/$^{16}$O ratio is, however, smaller by more than a factor of two than for the model calculations and for the Solar System. This small ratio cannot, under any circumstance, be caused by the simple mixing of WR material with solar-system abundances. It is likely that at least part of the explanation is elemental and mass fractionation of the GCR source material. Cassé & Goret (1978) recognized that elements with a low first-ionization potential (FIP) had a GCR source to solar-system abundance ratio that was significantly enhanced over those with a high-FIP. An alternative model (Epstein 1980; Cesarsky and Bibring 1981) noted that most of the elements with low-FIP for which GCR source abundances had been determined were refractory, while those with high-FIP were volatile, suggesting that the material of GCRs might preferentially originate in interstellar dust. For many years, the similarity between GCR source abundances and abundances in SEPs was taken as support for FIP being the governing property, rather than volatility. More recent work by Meyer et al. (1997) and Ellison et al. (1997) has given support to a model in which the GCR fractionation is governed by volatility. In this model the refractory elements are enriched in the GCRs since they sputter off accelerated dust grains, and are thus more easily accelerated by SN shocks.

Although atomic or molecular oxygen is highly volatile, nearly a quarter of the oxygen in the ISM is believed to exist in refractory compounds, e.g. in silicates (Lodders 2003). Thus in the Meyer et al. and Ellison et al. models, that fraction of the oxygen should be preferentially injected into the GCRs. On the other hand, a significant fraction of carbon, which is refractory in its elemental form, exists in the ISM as a volatile in molecules such as CO (Meyer et al. 1997). In addition, nitrogen exists primarily as a gas. So both the $^{12}$C/$^{16}$O and the $^{14}$N/$^{16}$O GCR ratios should be corrected for this effect to have a strictly valid comparison. We can make a rough adjustment to the $^{14}$N/$^{16}$O ratio since the fraction of $^{14}$N and $^{16}$O that exists in the solid state in the pre-solar nebula has been estimated. According to Lodders (2003), 23% of oxygen in the pre-solar nebula is in the solid state, and nearly all of the nitrogen is in the gaseous state. Meyer et al. (1997) show that the GCR source to solar-system abundance ratio for the refractory elements is roughly a factor of 13 larger than for nitrogen. They also point out that, even for volatile elements there appears to be a systematic enhancement in the abundance of heavy volatiles compared to light volatiles. They estimate the dependence of this enhancement on mass (A) as $A^{0.8\pm0.2}$. If we assume that the oxygen in grains is injected into the GCRs with an efficiency 13 times that of the fraction that is in the gas phase and, in addition, make an adjustment for the differing mass of $^{14}$N and $^{16}$O for the volatile oxygen fraction, then the $^{14}$N/$^{16}$O GCR source ratio should be increased by a factor $(0.23\times13)+0.77\times(16/14)^{0.8}=3.85$ to find the ratio for the source material prior to acceleration. This adjusted ratio is plotted as an open diamond in Figure 6. The error bar was obtained by scaling the unadjusted error bar by the ratio of the adjusted to the unadjusted $^{14}$N/$^{16}$O ratios, and then adding this error quadratically to the uncertainty resulting from the uncertainty in the mass dependence exponent. We have not included an uncertainty associated with the fraction of $^{16}$O that exists in the refractory and gas states. The $^{12}$C/$^{16}$O ratio is more difficult to correct since the fraction of carbon that is in the solid state in the ISM is poorly known. Therefore we have not attempted this adjustment.

We can reduce the effect of fractionation based on volatility if we look at the ratio of elements such as N/Ne that exist almost entirely in the volatile state in the ISM. In Figure 6 we have plotted the measured N/Ne ratio as a solid diamond and see that it is nearly 40% lower than for solar-system abundances. We have adjusted the N/Ne ratio for the mass dependent enhancement and plotted it as an open diamond in Figure 6. The error bar on the adjusted point was obtained using the same method as for $^{12}$C/$^{16}$O above. The $^{22}$Ne/$^{20}$Ne ratio has similarly been adjusted for mass dependence and the adjusted ratio is plotted as an open circle.

After these adjustments are applied the $^{14}$N/$^{16}$O and N/Ne are in much better agreement with both solar-system and the WR modeling results. However, the adjusted ratios should be regarded as approximate values showing that ratios previously thought to be inconsistent with solar-system abundances may very well be consistent if GCRs are fractionated on the basis of volatility and mass (Meyer et al. 1997; Ellison et al. 1997), and that fractionation is properly taken into account. Because of the model dependent nature of these adjustments, the values quoted throughout the paper for the $^{22}$Ne/$^{20}$Ne source ratio do not include this adjustment

Taken as a whole, we see that after adjustments for elemental fractionation, the CRIS data combined with those from other experiments show an isotopic composition similar to the one obtained by mixing about 20% of WR wind material with about 80% of material of solar-system composition. The largest ratios predicted by the WR models (including fractionation adjustments), $^{12}$C/$^{16}$O, $^{22}$Ne/$^{20}$Ne, and $^{58}$Fe/$^{56}$Fe are in fact observed. All other measured ratios are in reasonable agreement with small or insignificant differences from WR model predictions, which are very similar to solar-system abundances. We take this agreement as evidence, in addition to that already obtained from previous measurements of the $^{22}$Ne/$^{20}$Ne ratio (see references in section 1), that WR star ejecta is likely an important component of the cosmic-ray source material.

The WR models discussed above do not explicitly assume that the GCR origin is in superbubbles. However, the arguments made by Higdon and Lingenfelter (2003) that most WR stars reside in superbubbles, as do most core-collapse SNe, would appear to indicate that superbubbles are the predominant site of injection of WR material into the GCR source material. A clear corollary to this conclusion is that SN ejecta within the superbubble must also be accelerated by the same shocks that accelerate the WR ejecta. Therefore the picture that emerges from these data alone is that superbubbles would appear to be the site of origin and acceleration of at least a substantial fraction of GCRs.

The CRIS measurements of the $^{59}$Ni and $^{59}$Co isotopes (Wiedenbeck et al. 1999), which show that the $^{59}$Ni in GCRs has completely decayed, have led us to conclude that refractory GCRs must reside in an atomic or molecular state, most likely in interstellar grains (Ellison et al. 1997), for a time $>\sim 10^5$ years before acceleration to GCR energies, since $^{59}$Ni decays only by electron capture. (As nuclei are accelerated to GCR energies, the orbital electrons are quickly stripped off and nuclei that decay only by electron-capture become stable.)

Therefore, the $^{59}$Ni/$^{59}$Co results appear to be consistent with the Higdon et al. (1998) suggestion that GCRs are being accelerated from dust and gas within superbubbles. Dust has been observed around ~30% of all known WR stars in the WC phase (van der Hucht 2001), and some of these stars have been identified as belonging to OB associations (van der Hucht 2001; Williams et al. 2001; Knödlseder et al. 2002; Niedzielski 2003). So the scenario that is suggested is that WR star ejecta, enriched in $^{22}$Ne and some other neutron-rich isotopes, mixes with ejecta from core-collapse supernovae, and with average ISM (represented by solar-system abundances) in the tenuous medium within a superbubble. The refractory elements in this mix must exist mostly as grains and the volatiles primarily as gas. The mean time between SN events within superbubbles is estimated to be ~3-35 x $10^5$ years (Higdon & Lingenfelter 2003), providing sufficient time for $^{59}$Ni to decay to $^{59}$Co. Shocks from SNe within the superbubble, occurring on average on a time scale $>10^5$ years, then accelerate the mix of material in the superbubble to cosmic-ray energies, with the grains being preferentially accelerated according to the mechanism developed in detail by Ellison et al. (1997).

Recent discoveries of TeV γ-ray sources by the ground-based High Energy Gamma-Ray Astronomy (HEGRA) and High Energy Stereoscopic System (HESS) telescopes are a clear indication that cosmic-ray acceleration to high energies is occurring at those sites. Currently, a total of 15 TeV gamma-ray sources have been identified. A number of these have been shown to be spatially coincident with SNRs in our galaxy (Aharonian, et al. 2005a). Additionally, three of these sources are spatially coincident with OB associations. The source TeV J2032+4130 is spatially coincident with Cygnus OB2 (Aharonian et al. 2005b), and HESS J1303-631 and PSR B1259-63/SS2883 are spatially coincident with Cen OB1 (Aharonian et al. 2005c). Furthermore, the Wolf-Rayet star θ-Mus is a member of this OB association (Aharonian et al. 2005c). In addition, HESS J1804-216 coincides spatially with SNR G8.7-0.1 "which is known to be associated with molecular gas where massive star formation is taking place" (Aharonian, et al. 2005a). These discoveries of TeV γ-ray sources, some of which are spatially coincident with OB associations, strengthen our conclusion obtained from galactic cosmic rays at much lower energies, that superbubbles are the source of at least a substantial fraction of galactic cosmic rays.

Additional work in comparing our cosmic-ray data with superbubble and WR models by varying model parameters is clearly needed. For example, it would be of interest to compare the GCR results with WR model predictions for metallicities other than Z=0.02 (believed today to be Z=0.0122); Asplund et al. 2004) to see how the comparison between experimental results and these model calculations change. It would also be useful to explore the model sensitivity to uncertainties in nuclear reaction cross-sections and to include the SN ejecta along with WR ejecta.

Models dominated by a high initial-mass, rotating WR component, are clearly excluded. Initial mass functions typically used to describe the mass distribution of massive stars (Salpeter 1955) predict a rapidly decreasing number of stars with increasing mass, so we would not expect high mass stars to dominate in any case.

## 5. SUMMARY

Our measurements have led to an improved value for the $^{22}$Ne/$^{20}$Ne source abundance ratio that is a factor of 5.3±0.3 greater than for the solar wind. This ratio is significantly larger than in any other known sample of "cosmic" matter with the exception of meteoritic Neon-E. A comparison of measurements from CRIS and from other experiments with stellar model predictions shows that for non-rotating and M<85 $M_\odot$ rotating WR models, the three isotope ratios predicted to be most enhanced relative to the solar system, $^{12}$C/$^{16}$O, $^{22}$Ne/$^{20}$Ne, and $^{58}$Fe/$^{56}$Fe, are indeed present in the GCRs. All other measured ratios are in reasonable agreement with small or insignificant differences from WR model predictions, which are very similar to solar-system abundances, provided that elemental ratios in GCR source abundances are fractionated according to the volatility model of Meyer et al. (1997).

We take this agreement as evidence, in addition to that previously suggested by earlier measurements of the $^{22}$Ne/$^{20}$Ne ratio, that WR star ejecta is likely an important component of the cosmic-ray source material. Since most WR stars reside in superbubbles, as do most core-collapse supernovae, superbubbles must be the predominant site of injection of WR material into the GCR source material. Therefore the picture that emerges from these data is that superbubbles would appear to be the site of origin and acceleration of at least a substantial fraction of GCRs.

## APPENDIX A

The cross-sections used in the "tracer method" propagation described in section 3 were derived from cross-section measurements taken from the literature for the most important cross-sections contributing to our estimate of the $^{22}$Ne/$^{20}$Ne ratio, i.e. those for $^{24}$Mg and $^{28}$Si projectiles fragmenting into the neon isotopes and the fluorine and oxygen tracer isotopes. These measurements were obtained using energetic proton projectiles on Mg and Si targets. All other cross-sections were taken from Silberberg et al. (1998), scaled to measured cross-sections obtained using beams of energetic heavy ions when they were available. In our previous work (Binns et al. 2001), all cross-sections were taken from Silberberg et al. (1998), scaled to available cross-section measurements. There were two classes of measurements from which our cross-sections were derived. The first class consists of direct measurements of undecayed cross-sections obtained at accelerators using electronic pulse instruments. The second class was cumulative, or decayed, measurements obtained at accelerators using x- and γ-spectrometry, and accelerator mass spectrometry (AMS). There are two significant differences in the measurements for these two data classes. The first is that the cumulative measurements contain not only particles that fragment directly into the isotope reaction product of interest, but also those isotopes that β-decay into that isotope in a time short compared to the time scale between exposure and analysis (typically several weeks). This is in contrast to direct measurements that are made on a time scale short compared to β-decay half-lives. The second is that for the cumulative measurements protons are projected onto targets of Mg and Si that contain natural abundances of their isotopes. Thus, to obtain the cross-section for a particular isotope on hydrogen, it is necessary to correct the cross-section that was measured using the natural target.

In Figures A1 and A2 we plot the cumulative (filled circles) and the direct (open circles) cross-section measurements for Mg and Si projectiles respectively. In addition, for comparison we have plotted curves derived from Silberberg et al. (1998) cross-sections. The energies of cosmic rays contributing fragments relevant for the CRIS instrument range from a few-hundred MeV/nucleon to ~1

GeV/nucleon. Over this energy range, for most of the reactions, it is difficult to see any pronounced energy dependence. Therefore we have taken the cross-section for these reactions to be energy independent. The direct and cumulative measurements were treated separately and combined at the end of the calculation. "Best value" cross-sections and uncertainties were calculated using the following method.

The weighted mean, the uncertainty of the weighted mean, and the reduced $X^2$ were calculated for each reaction using only the measurements over the energy range 380-1200 MeV/nuc. If $X^2_{red}$ <1, then, to avoid having the weighted mean being dominated by measurements reported with very small error bars, each of the measurement uncertainties was broadened by adding in quadrature the arithmetic mean of the measurements multiplied by a constant. The value of this constant was then adjusted using these modified uncertainties until the $X^2_{red}$ =1. This adjusted uncertainty was then taken as the uncertainty for that data point and a new adjusted weighted mean and uncertainty of that weighted mean was calculated using the modified uncertainties.

For the direct measurements, this adjusted weighted mean and uncertainty were the values used. However, as mentioned above, the cumulative measurements were obtained using Mg and Si targets with natural abundances. Therefore, to obtain the cross-section for the interaction of the dominant isotope projectiles (i.e. $^{24}$Mg and $^{28}$Si) in those measurements, it was necessary to correct the cross-sections measured using the natural target. The cross-sections were corrected for this effect by multiplying the cumulative cross-sections for each projectile isotope times its natural abundance fraction (f's in equation below) and setting the sum equal to the measured cumulative cross-section. For example, the equation below was used to obtain the cumulative cross-section for $^{24}$Mg fragmenting into $^{21}$Ne.

$$\sigma^{cum\ meas}_{Mg} = \sigma^{cum}_{12,24 \to 10,21} \times f_{12,24}$$
$$+ (\sigma^{S\&T}_{12,25 \to 11,21} + \sigma^{S\&T}_{12,25 \to 10,21} + \sigma^{S\&T}_{12,25 \to 9,21}) \times f_{12,25}$$
$$+ (\sigma^{S\&T}_{12,26 \to 11,21} + \sigma^{S\&T}_{12,26 \to 10,21} + \sigma^{S\&T}_{12,26 \to 9,21}) \times f_{12,26}$$

The measured cross-section in the natural material is $\sigma^{cum\ meas}_{Mg}$. The cross-sections for the target isotopes with small abundance ($^{25}$Mg and $^{26}$Mg) were taken to be the Silberberg et al (1998) cross-sections, averaged over the energy interval 380-1200 MeV/nuc. Since the natural abundance fraction for each isotope is known, the equation was then solved to obtain the single-isotope, cumulative cross-section $\sigma^{cum}_{12,24 \to 10,21}$. To obtain the uncertainty assigned to this cross-section, the adjusted uncertainty of the weighted mean described above was used for the cumulative measured uncertainty. The Silberberg et al. cross-sections were arbitrarily assigned an uncertainty of 50% of the cross-section value and the overall cumulative uncertainty was then calculated.

The direct and cumulative cross-section weighted means and uncertainties for each reaction were then combined to give their total weighted means and uncertainties. Table A1 gives the final cross-sections and uncertainties derived using this method, which were used in our propagation. These are also shown in Figure A1 and A2 as solid horizontal bars. Note that in some cases the adopted value for the cross-sections for a given reaction fall above or below all of the data points. This is a result of the adjustment required to obtain the interaction cross-section for a single isotope from measurements made using targets with natural isotope abundances for that element.

| Fragment (Z,A) | $^{24}$Mg cross-section (mb) | Uncertainty (mb) | $^{28}$Si cross-section (mb) | Uncertainty (mb) |
|---|---|---|---|---|
| 10,22 | 48.2 | 1.5 | 23.7 | 0.7 |
| 10,21 | 29.7 | 1.1 | 21.5 | 1.1 |
| 10,20 | 30.6 | 1.0 | 21.5 | 1.0 |
| 9,19 | 16.1 | 1.5 | 12.6 | 1.0 |
| 8,17 | 15.9 | 2.0 | 12.0 | 2.3 |

Table A1

Table of derived cross-sections and uncertainties for $^{24}$Mg and $^{28}$Si fragmenting into the isotopes listed in column 1.

*Acknowledgments*

We wish to thank Katharina Lodders of Washington University for helpful discussions on the fraction of elements tied up in grains in the Solar System interplanetary medium. We also wish to thank Sachiko Amari, Charles Hohenberg, and Ernst Zinner for helpful discussions on neon in meteorites. This research was supported by the National Aeronautics and Space Administration at the California Institute of Technology (under grants NAG5-6912 and NAG5-12929), Washington University, the Jet Propulsion Laboratory, and the Goddard Space Flight Center.

Yanasak, N. E., et al. 2001, ApJ, 563, 768

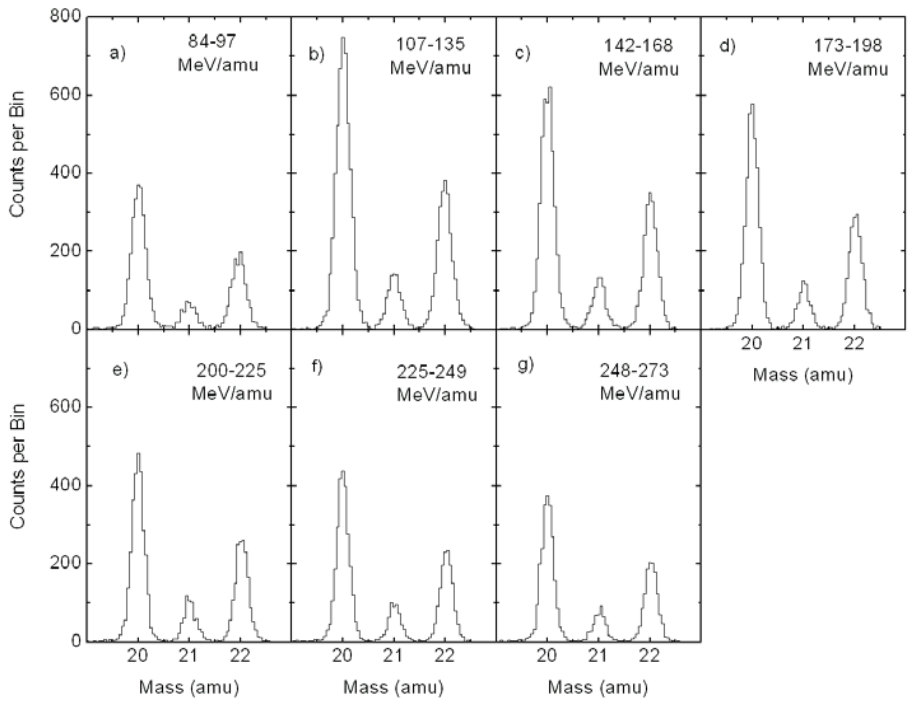

Figure 1—Mass histograms of neon events stopping in each of seven layers of silicon detectors. The energies listed in the figures are for $^{20}$Ne. The corresponding energies of $^{21}$Ne and $^{22}$Ne are slightly lower.

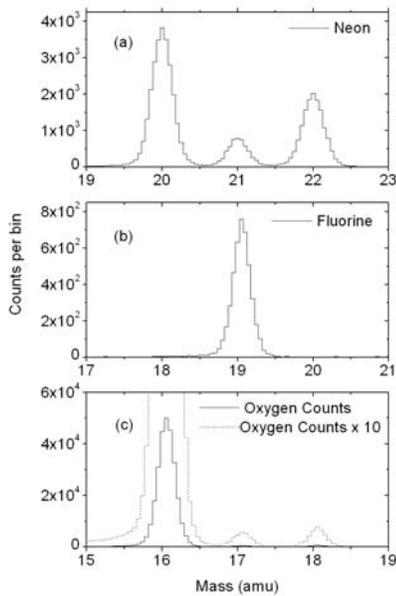

Figure 2—Mass histograms summed over the 7 ranges shown in Figure 1 for (a) neon, (b) fluorine, and (c) oxygen.

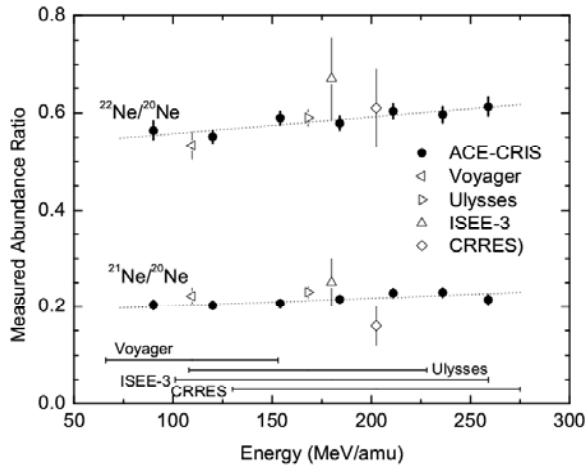

Figure 3—The ACE-CRIS measurements of the ratios $^{22}$Ne/$^{20}$Ne and $^{21}$Ne/$^{20}$Ne are plotted as a function of energy. Abundances measured by other experiments (Wiedenbeck & Greiner 1981 [ISEE-3]; Lukasiak et al. 1994 [Voyager]; Connell & Simpson 1997 [Ulysses]; DuVernois et al. 1996 [CRRES]) are plotted as open symbols and the energy intervals for their measurements are shown as horizontal bars at the bottom of the figure.

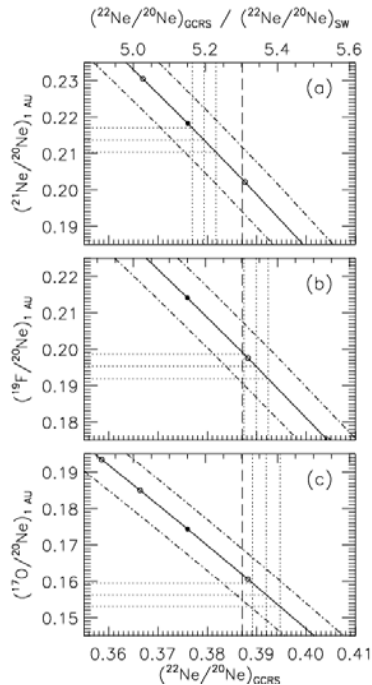

Figure 4—Source abundance of the $^{22}$Ne/$^{20}$Ne ratio calculated using the tracer isotopes a) $^{21}$Ne, b) $^{19}$F, and c) $^{17}$O. The solid diagonal line shows how the inferred source abundance ratio of $^{22}$Ne/$^{20}$Ne and the ratio of the tracer isotope to $^{20}$Ne vary with escape mean free path. The center horizontal dotted lines for each tracer isotope are the measured ratio of that tracer to $^{20}$Ne, and the top and bottom lines are the corresponding 1-σ measurement uncertainties. The intersection of the center vertical dotted line is the best estimate of the GCR source $^{22}$Ne/$^{20}$Ne ratio resulting from that tracer isotope. The left and right dotted lines are the corresponding 1-σ measurement uncertainties due to the uncertainty in the ratio of the tracer isotope to $^{20}$Ne. The vertical dashed line is the weighted mean of the three tracer ratios.

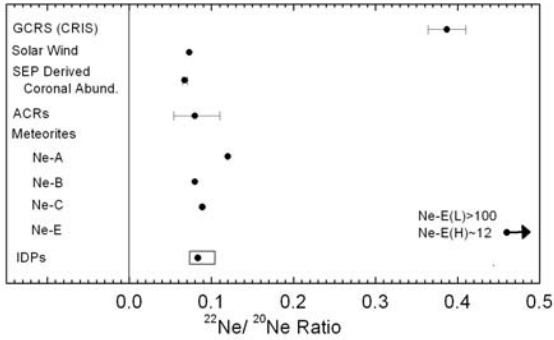

Figure 5—The CRIS $^{22}$Ne/$^{20}$Ne source abundance for GCRs is compared to solar wind (Geiss,1973; Anders and Grevesse, 1989), SEP derived coronal abundances (Leske et al. 2003), ACR's (Leske et al. 1996 and 1999a), meteoritic abundances (Ozima and Podosek, 1983), and IDP's (Kehm, 2000). The plotted error bar for CRIS is the quadratic sum of the statistical and systematic uncertainties (see text). The data point for SEPs and its uncertainty indicate the value deduced for the SEP source after accounting for the fractionation. The point plotted for IDPs is the average value obtained for 27 of the particles measured, and the horizontal bar indicates the spread of these measurements.

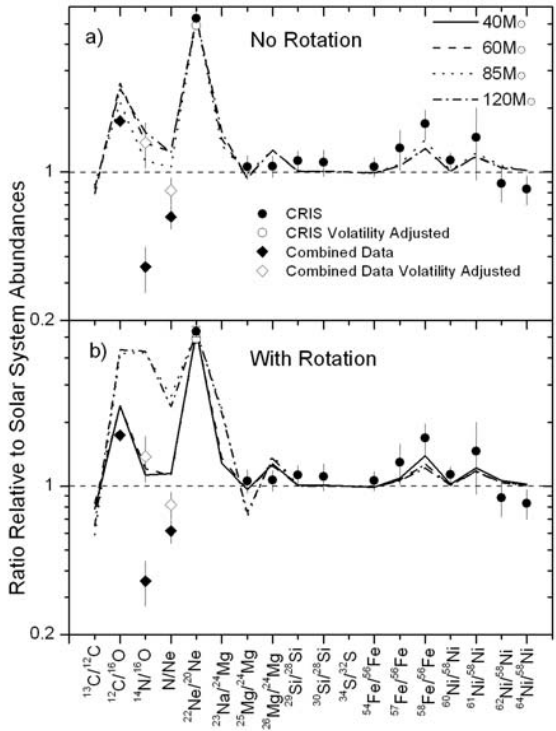

Figure 6—CRIS ratios compared with model predictions for WR stars with (6a) no rotation, and (6b) an equatorial surface rotation velocity of 300 km s$^{-1}$ for the initial precursor star for masses of 40, 60, 85, and 120 M$_\odot$, and for metallicity Z$_\odot$=0.02. The plotted neon, magnesium, silicon, iron, and nickel source abundance ratios are from CRIS data (Wiedenbeck et al. 2001a, 2001b, & 2001c). The closed diamonds plotted are mean values of ratios, weighted by their published uncertainties, obtained from Ulysses (Connell and Simpson, 1997), ISEE-3 (Krombel and Wiedenbeck, 1988; Wiedenbeck and Greiner, 1981), Voyager (Lukasiak et al. 1994) and HEAO-C2 (Engelmann et al. 1990). The plotted mean values are obtained from these experiments as follows: $^{12}$C/$^{16}$O—Ulysses and HEAO-C2 (these are actually element ratios that have not been corrected for the small fraction of neutron rich C and O isotopes present at the source); $^{14}$N/$^{16}$O—ISEE-3, Voyager, and HEAO-C2; N/Ne—Ulysses and HEAO-C2. The open diamonds are the $^{14}$N/$^{16}$O and N/Ne ratios adjusted for a volatility fractionation.

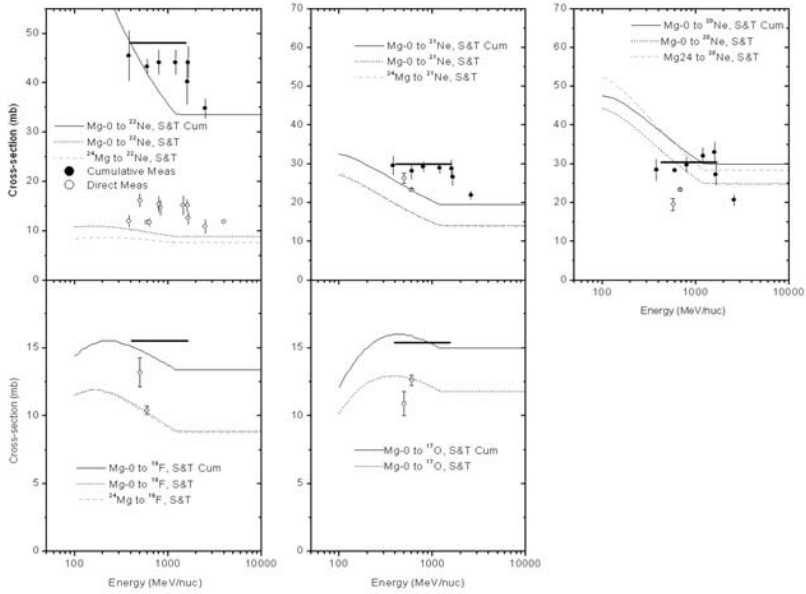

Figure A1—Measured cross-sections in millibarns for magnesium used to derive the "best value" cross-sections for our propagations are plotted. The cumulative measurements (closed circles) were obtained using protons incident upon a magnesium target with natural abundances and were taken from Michel et al. 1989, 1995, and 1996, Schiekel et al. 1996, and Leya et al., 1998. The direct measurements (open circles) were obtained using a $^{24}$Mg projectile incident upon a hydrogen or polyethylene and carbon targets and were taken from Webber et al. 1990, 1998a, and 1998b. Additionally, direct cross-sections for $^{22}$Na were taken from Michel et al. 1989, Michel et al. 1989, and Leya et al. 1998. The curves are derived using the Silberberg et al. 1998 cross-sections. They are: 1) solid line--the cumulative cross-section for protons fragmenting on a natural Mg target, 2) dashed line--the direct cross-section for protons fragmenting on a natural Mg target, and 3) dotted line--the direct cross-section between specific projectile and fragment isotopes. Mg-0 refers to a natural abundance magnesium target. The solid horizontal line gives the final cross-sections and uncertainties derived using this method, which are listed in Table A1.

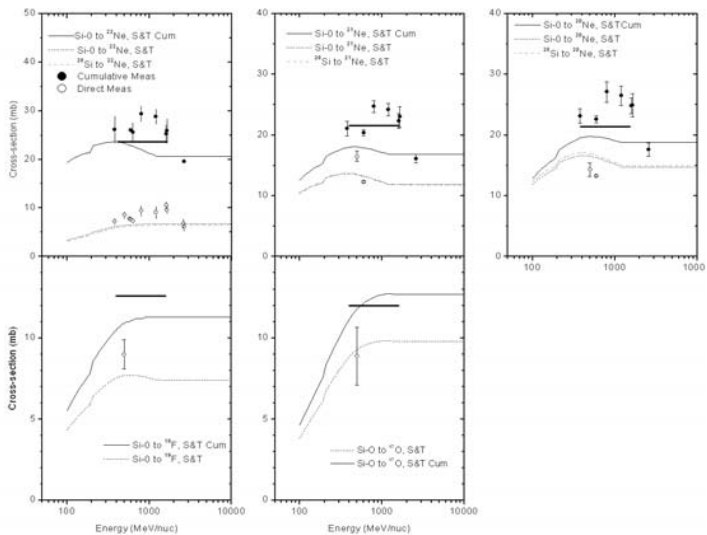

Figure A2—Same as A1a but for silicon.